\documentclass[twocolumn,superscriptaddress,amsmath,amssymb,aps,prl,reprint,floatfix]{revtex4-1}

\usepackage{graphicx}
\usepackage{dcolumn}
\usepackage{bm}
\usepackage{float}

\newcommand{\matrixel}[3]{\ensuremath{\left\langle #1 \middle| #2 \middle| #3 \right\rangle}}

\begin{document}

\title{Ab-initio simulations and measurements of the free-free opacity in Aluminum}

\author{P.~Hollebon}
\email{patrick.hollebon@physics.ox.ac.uk}
\affiliation{Department of Physics, Clarendon Laboratory, University of Oxford, Parks Road, Oxford OX1 3PU, UK}
\author{O.~Ciricosta}
\affiliation{Department of Physics, Clarendon Laboratory, University of Oxford, Parks Road, Oxford OX1 3PU, UK}
\author{M.P.~Desjarlais}
\affiliation{Pulsed Power Sciences Center, Sandia National Laboratories, Albuquerque, NM 87185, USA}
\author{C.~Cacho}
\author{C.~Spindloe}
\author{E.~Springate}
\author{I.C.E.~Turcu}
\affiliation{Central Laser Facility, STFC Rutherford Appleton Laboratory, Didcot OX11 0QX, UK}
\author{J.S.~Wark}
\affiliation{Department of Physics, Clarendon Laboratory, University of Oxford, Parks Road, Oxford OX1 3PU, UK}
\author{S.M.~Vinko}
\email{sam.vinko@physics.ox.ac.uk}
\affiliation{Department of Physics, Clarendon Laboratory, University of Oxford, Parks Road, Oxford OX1 3PU, UK}

\date{\today}

\begin{abstract}

The free-free opacity in dense systems is a property that both tests our fundamental understanding of correlated many-body systems, and is needed to understand the radiative properties of high energy-density plasmas. Despite its importance, predictive calculations of the free-free opacity remain challenging even in the condensed matter phase for simple metals. Here we show how the free-free opacity can be modelled at finite-temperatures via time-dependent density functional theory, and illustrate the importance of including local field corrections, core polarization and self-energy corrections. Our calculations for ground-state Al are shown to agree well with experimental opacity measurements performed on the Artemis laser facility across a wide range of x-ray to ultraviolet wavelengths. We extend our calculations across the melt to the warm-dense matter regime, and find good agreement with advanced plasma models based on inverse bremsstrahlung at temperatures above 10\,eV.

\end{abstract}


\maketitle

The mechanisms by which free electrons in a plasma absorb and emit radiation are of key importance to a range of applications, from investigations of laser-plasma interactions to astrophysics and inertial confinement fusion research.
The free-free opacity in classical plasmas is generally described using the inverse bremsstrahlung (IB) formalism, initially treated classically by Kramers\,\cite{Kramers1923}, and later modified to include a range of additional corrections to the absorption cross section including quantum effects\,\cite{Johnston1967}, multi-photon contributions\,\cite{Seely1973}, relativistic corrections\,\cite{Tsytovich1995}, electron degeneracy\,\cite{Totsuji1985} and collective phenomena\,\cite{Dawson1962,Bekefi1966,Ron1963}. These models contain Coulomb logarithm terms to describe electron interactions and are typically limited in applicability to plasmas where small-angle collisions dominate energy transfer in the electron subsystem, i.e., plasmas at high temperatures and low densities.

Dense plasmas, in turn, have proven far more challenging both to model and to investigate experimentally\,\cite{Vinko2009a,Iglesias2010,Williams2013,Kettle2016,Shaffer2017,Williams2018}. Perhaps surprisingly, similar difficulties are encountered in condensed matter systems such as ground-state and liquid metals. Here, the absorption process is generally treated using linear response theory via calculations of the dielectric function or the complex conductivity\,\cite{Hopfield1965,Sturm1973}. Because the dielectric function provides a full description of the system's response, opacity investigations can be used to validate {\it ab-initio} models more generally, and provide stringent constraints on the approximations deployed to study correlated many-body systems~\cite{Sturm1982,Sturm1990,Zylstra2015,Roth2017,Baczewski2016}. In particular, simple metals irradiated at photon energies below their bound-state ionization edges are ideal candidates to investigate the fundamental mechanisms of free-free opacity, and therefore of the dielectric response. Such systems are essentially ground-state plasmas, both simple to manipulate and investigate experimentally, and are present in well-defined conditions of temperature, density and ionization.

Despite this, both the theoretical and the experimental free-free absorption cross sections in Al, a simple metal, remain poorly understood.
While calculations based on the random phase approximation (RPA) are seen to be in good agreement with bound-state opacity measurements, the RPA performs poorly in the free-free regime even for ground-state Al~\cite{Sturm1990}. Furthermore, there is a significant discrepancy in the experimental free-free opacity in the XUV photon energy range between the Al plasma frequency at 15\,eV and the Al L-edge at 73\,eV. The widely-used Centre for X-Ray Optics online database (CXRO)\,\cite{CXRO} uses the experimental data of Gullikson et al.\,\cite{Gullikson1994} for this energy range, results which disagree by as much as a factor of two with measurements by Henke et al.\,\cite{Henke1993} and Keenan et al.\,\cite{Keenan2002}. Given that thin Al foils are commonly used as filters in the XUV, such discrepancies can have a large effect on calculated spectral brightnesses of XUV sources created via high harmonic generation (HHG) or other techniques.

Here we present experimental measurements of the ground-state Al free-free opacity, which are seen to be in excellent agreement with theoretical calculations based on finite-temperature time-dependent density functional theory (DFT).
We show that it is necessary to go beyond the Kubo-Greenwood approximation in DFT to model the free-free opacity, and illustrate the contributions of both local field corrections (LFC) and G$_{0}$W$_{0}$ quasi-particle corrections. We extend our calculations to finite-temperature equilibrated warm-dense matter up to temperatures of 15\,eV, and find our calculations approach IB-based plasma models at higher temperatures.

\begin{figure}
\includegraphics[width=\linewidth]{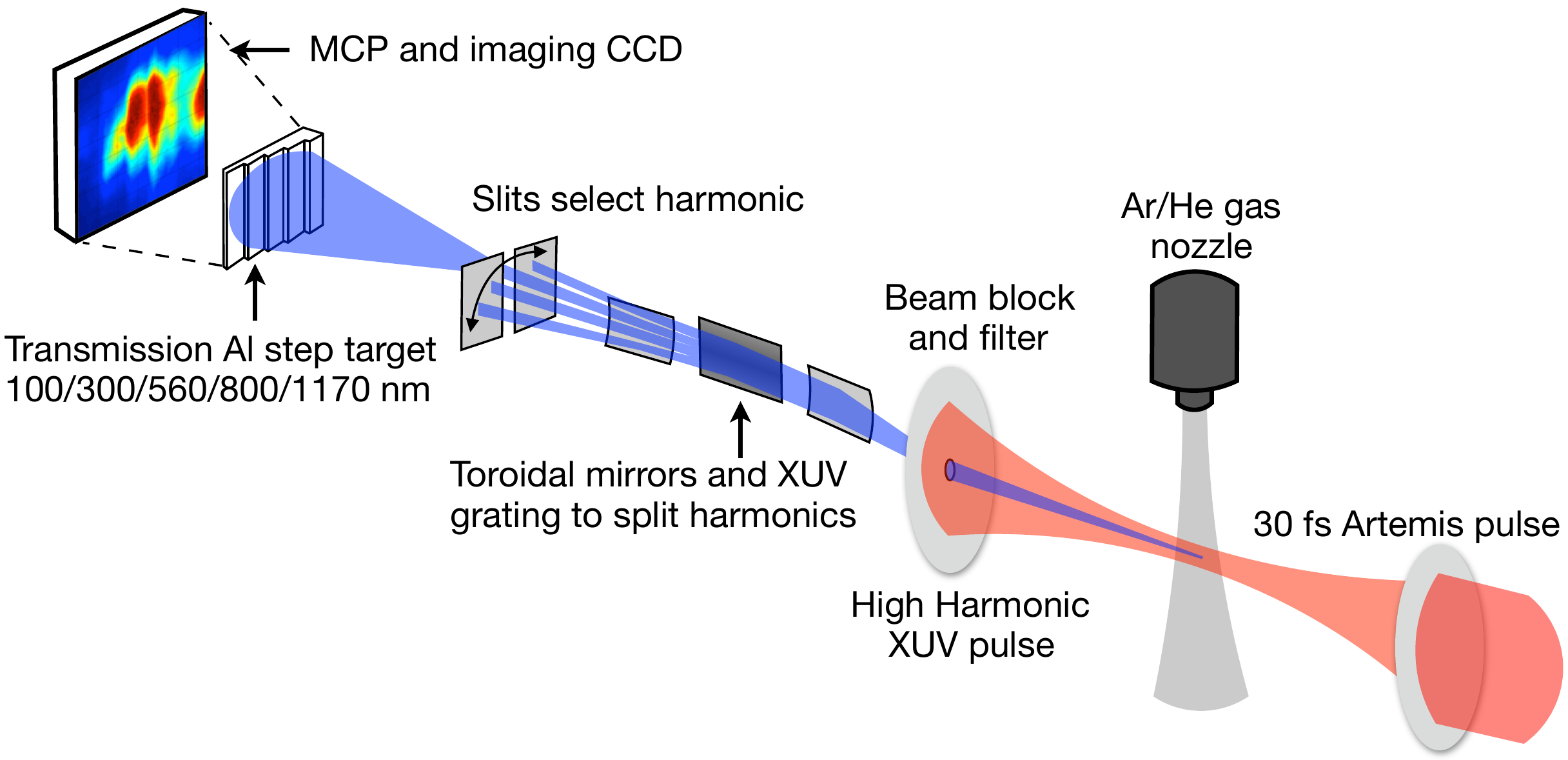}
\caption{(Color online). Experimental setup for the opacity measurement on the Artemis facility.}
\label{FIG:setup}
\end{figure}

The experiment was conducted at the HHG Artemis facility~\cite{Frassetto2011} of the Rutherford Appleton Laboratory, UK, capable of delivering XUV pulses in the XUV spectral range between 15-60\,eV. The setup is shown schematically in Fig.\,\ref{FIG:setup}. Here, a 1 kHz Ti:sapphire CPA laser system provided short (30\,fs FWHM) infrared pulses at a wavelength of 780 nm. These pulses were focused onto an Ar or Ne gas jet within a differentially pumped gas cell at intensities around 10$^{14}$\,W/cm$^2$ to produce high harmonics over a relatively broad spectrum, with an energy efficiency of order 10$^{-6}$. Individual harmonics are picked out from the HHG spectrum via a time-preserving XUV monochromator with a resolving power $\lambda/\Delta \lambda \approx$100 and peak transmissivity of 20\%\,\cite{Frassetto2011}, and are steered onto an Al target using a gold mirror. The transmitted beam through the target hits a microchannel plate detector and is imaged with a standard optical system and camera. The background signal and spatial profile of the HHG pulses was determined by imaging the beam in the absence of the sample. The linearity of the system with respect to the HHG signal was validated experimentally.

The target sample is a free-standing five-step foil with the different thicknesses of Al deposited in a single process to ensure there are no layers of spurious materials between the individual steps. The target size is about 3\,mm by 4\,mm, and is fully illuminated by the diverging HHG beam from the source (Fig.\,\ref{FIG:setup}). Thicknesses of the various steps vary slightly with position, but are on average 100\,nm, 315\,nm, 560\,nm, 810\,nm and 1165\,nm, measured via profilometry to an accuracy better than 2\,nm. 
Using a single Al sample with a range of thicknesses is crucial because surface contaminants and oxide layers can have a large effect on the absorption, given the vastly different attenuation lengths in the XUV regime. By looking at the differential absorption across the steps we can overcome the well-known difficulties related to surface oxide layers and other contaminants commonly present on thin foils. The absorption of XUV light in the low-intensity limit follows the Beer-Lambert law:
\begin{equation}
T(x,y) = \exp(-\kappa(\omega) d(x,y) - \alpha(\omega)).
\end{equation}
Here, $T$ is the experimentally measured transmission through a specific point of the target denominated by $(x,y)$. The thickness of the absorbing Al target at that point is $d(x,y)$, $\kappa(\omega)$ is the frequency-dependent absorption coefficient of Al, and $\alpha(\omega)$ is the frequency-dependent absorption term (including thickness) corresponding to any additional contribution, primarily dominated by aluminum oxide. We implicitly assume that there is no variation in oxide thickness across the target. By scanning across different values of $d$ of the target at a fixed photon energy ($\hbar\omega$) we obtain a range of values for the transmission $T$ and can fit the measurement to determine both $\kappa(\omega)$ and $\alpha(\omega)$ independently. To obtain the correct absorption coefficient, it is therefore only necessary to know the difference in thicknesses of the various steps, while their absolute values, or the absorption by potential oxide layers, are inconsequential. 

\begin{figure}
\centering
\includegraphics[clip=false, width=1\linewidth]{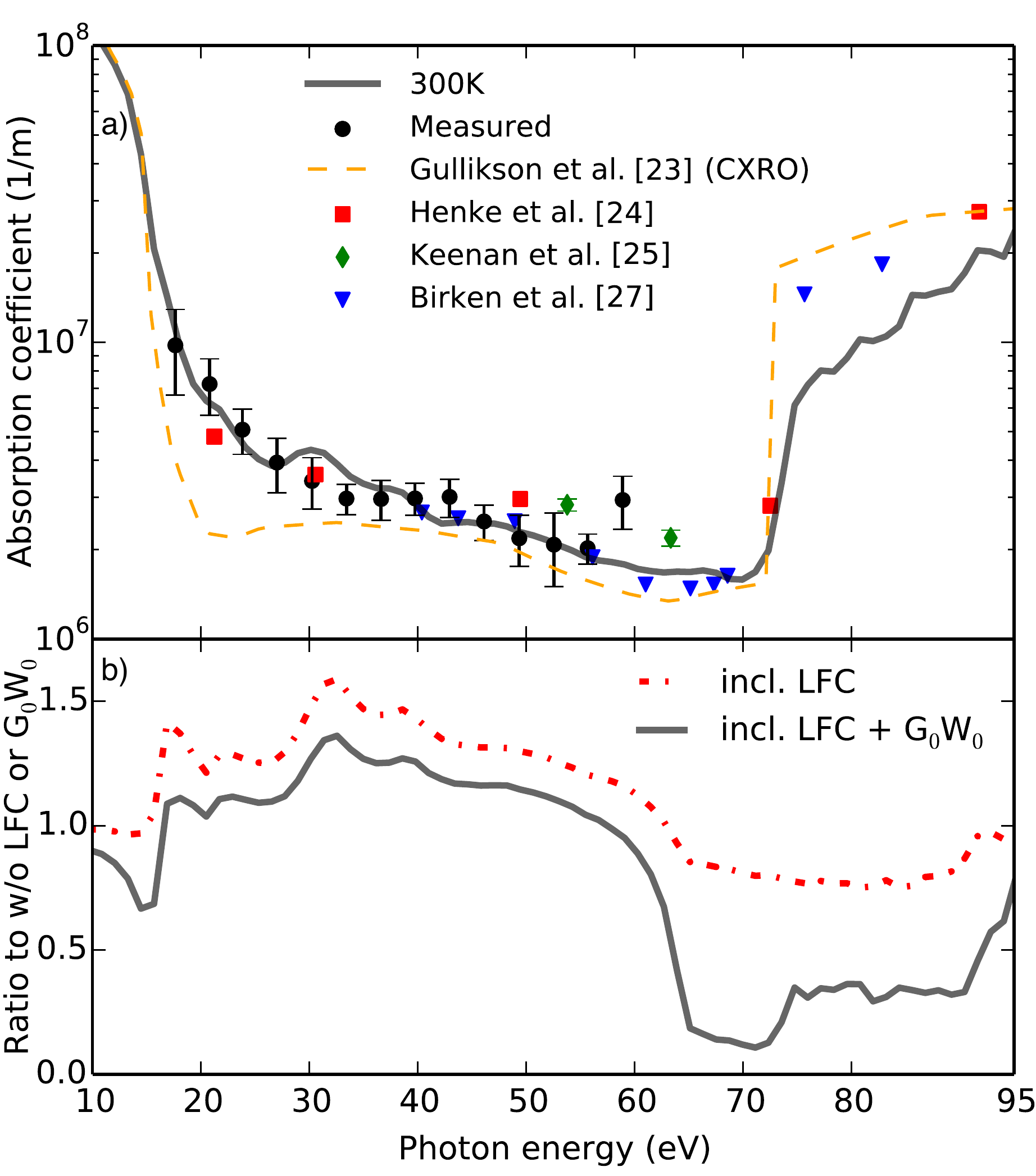}
\caption{(Color online). (a) Calculated opacity for room temperature Al along with our own measurements and those of previous studies. (b) LFC increases absorption between the plasma frequency and the L-edge, whilst the primary effect of the G$_{0}$W$_{0}$ correction is to shift the energy of the L-shell reducing the opacity. }
\label{fig:cold}
\end{figure}

We plot our experimental results alongside previous measurements, and our DFT-based calculations further described below, in Fig.\,\ref{fig:cold}a. We note good agreement between our measurements and those of Henke et al.\,\cite{Henke1993}, and Birken et al.\,\cite{Birken1986}, but a clear disagreement with the recently reported measurement of Kettle et al.\,\cite{Kettle2016}, and with the data by Gullikson et al.\,\cite{Gullikson1994} (CXRO) at photon energies below $\sim$30-40\,eV.
Our results also appear consistent with the recently reported cold opacity measurements by Williams et al.\,\cite{Williams2018} ($\kappa=2.5\times10^{6}$ m$^{-1}$), though as the latter measurement was not frequency resolved a more complete comparison is not possible.
If we assume that the dominant contribution to the offset coefficient $\alpha(\omega)$ is a layer of Al$_2$O$_3$ of known opacity, for example taken from the CXRO database\,\cite{CXRO}, then we can use the experimental data to also deduce the oxide thickness. Following this approach we find the thickness of the total oxide layer to be (15$\pm$6)\,nm.

Our experimental measurements are compared with detailed {\it ab-initio} calculations based on DFT. DFT provides an in-principle exact means of solving the quantum mechanical, many-body Coulomb system. However, calculated quantities are limited by the range of validity of available exchange-correlation functionals. This is particularly relevant when dealing with dynamic properties of systems at finite-temperatures\,\cite{Baczewski2016,Witte2017}, which require explicitly time\,\cite{Maitra2016,VanLeeuwen1999,Maitra2001} and temperature dependent functionals\,\cite{Dharma-wardana2016,Perrot1984,Dharma-Wardana1981a,Perrot2000,Pribram-Jones2016}. An alternative approach is to tackle Hedin's equations\,\cite{Hedin1965,Hedin1999} for the interacting Green's functions within the GW approximation for the single-particle self energy operator, either iterated to consistency or as a `one shot' G$_{0}$W$_{0}$ correction to inputted, for example Kohn-Sham, wavefunctions. Both are standard approaches to the band gap problem in the ground state limit, but here we apply them at finite-temperatures \cite{Faleev2006} allowing us to investigate warm dense matter systems within the DFT formalism. GW and other approximations for the self energy have, however, previously been applied to plasma physics independently of DFT~\cite{Zimmermann1978,Ropke1978,Kraeft1986, Gunter1991,Seidel1995,Kremp2005,Lin2017a,Kas2017}. 

Our calculations consist of three steps. First, a multi-centred Kohn-Sham DFT\,\cite{W1964,Mermin1965,W1965} simulation is performed using Projector Augmented Wave (PAW)\,\cite{Blochl1994, Torrent2008} pseudopotentials with 1s electrons frozen in the core. Up to 1500 bands are calculated in 32 atom unit cells with a $2\times2\times2$ k-point grid Brillouin zone sampling and a planewave cutoff of 400\,eV. An ensemble of ion positions is obtained by evolving the system in time for 3\,ps in the Born-Oppenheimer approximation whilst coupled to a Nos\'{e}-Hoover thermostat. Second, we use Time-Dependent DFT (TD-DFT)\,\cite{Runge1984,Pribram-Jones2016} for the first order density response function within the RPA, and calculate the dielectric function of the system\,\cite{Martin2010} for select ion spatial configurations from the last 1\,ps of evolution. Finally, we use the the calculated dielectric function to compute the first order, finite-temperature G$_{0}$W$_{0}$ corrections to the Kohn-Sham energy levels for bound L-shell states and low lying continuum states, thereby addressing the infamous band gap problem\,\cite{Onida2002}. Here, this step corrects the position of the L-edge. The G$_{0}$W$_{0}$ corrected quasi-particle energies and states are then used to compute the final dielectric function. Whilst GW is frequently used in the ground state to correct band gaps to sub-eV accuracy, a detailed study of finite-temperature many-body effects on the structure within the continuum is beyond the scope of this paper. Our calculations for room-temperature Al are shown in Fig.\,\ref{fig:cold}a alongside several experimental measurements\,\cite{Birken1986,Henke1993, Gullikson1994,Keenan2002}. The full calculation including LFC and G$_{0}$W$_{0}$ agrees well with our measured free-free opacity and the experimental position of the L-edge. Our calculations appear to underestimate the opacity immediately above the L-edge, but show clear signs of approaching the CXRO database values at higher photon energies.

The inclusion of LFC stems from the definition of the macroscopic dielectric function $\epsilon_{M}(\omega)$ from which we obtain the opacity $\kappa=\frac{2\omega}{c}\text{Im}\sqrt{\epsilon_{M}(\omega)}$. In a periodic system we can write the dielectric function as $\epsilon_{GG'}(q,\omega)$, where $G$ and $G'$ are reciprocal lattice vectors, and $q$ lies within the first Brillouin zone. One has:
\begin{equation}
\epsilon_{M}(\omega)= \frac{1}{[\epsilon(q \rightarrow 0,  \omega)^{-1}]_{00}} \neq \epsilon(q \rightarrow 0,  \omega)_{00} .
\end{equation} 
Here the inequality is a result of the off-diagonal elements $\epsilon_{0G}$ and $\epsilon_{G0}$. The right-hand side of the inequality is the dielectric function in the long wavelength limit \emph{without} LFC, as would be obtained using the Kubo-Greenwood expression for the dynamic conductivity, commonly used in conjunction with {\it ab-initio} DFT calculations\,\cite{Desjarlais2002,Mazevet2005,Mazevet2005a,Witte2017}. Only in the case of a homogenous system do the LFC vanish. The contribution of the LFC and of G$_{0}$W$_{0}$ on the room-temperature opacity calculations is shown in Fig.\,\ref{fig:cold}b. The LFC overall raises the opacity between the plasma frequency and the L-edge, while the effect of the G$_{0}$W$_{0}$ correction is to shift the L-edge to the correct L-shell binding energy, and to slightly decrease the opacity.

We have used the code Abinit\,\cite{AbinitWebsite,Gonze2016a} to perform our DFT calculations and to compute the dielectric function\,\cite{Miyake2000,Shishkin2006}. The long wavelength limit intraband contributions to the pole at $\epsilon(q\rightarrow0,\omega\rightarrow0)_{00}$ were computed by fitting the dielectric function at finite wavevector using the single plasmon-pole approximation:
\begin{equation}
\epsilon^{-1}(q,\omega) = 1 + \frac{(\omega^{0}_{p})^{2}}{\omega(\omega + \text{i}\nu^{q} ) - (\omega_{p}^{q})^2},
\end{equation}
where the plasma frequency $\omega_{p}^{q}$ and broadening $\nu^{q}$ are q-dependent. We found the dielectric response at low frequencies to be well fitted by this functional form for all conditions studied. The $q\rightarrow0$ plasma frequency, and thus the pole at $\omega=0$, was then obtained by fitting for $\omega_{p}^{0}$ as well as by extrapolating our calculated Bohm-Gross relation for $\omega_{p}^{q}$ to $q=0$. The same values were obtained by both methods. 

The primary computational hurdle is the calculation of the correlation energy $\Sigma^{C}(\omega)$, with which we can define the self-energy $\Sigma=\Sigma^{X}+\Sigma^{C}$, where $\Sigma^{X}$ is the finite-temperature exchange term. This then features in the effective Schrodinger equation for quasi-particle states $\psi^{QP}$:
\begin{equation}
\left[ \frac{-\hbar^{2}}{2m_{e}} \nabla^{2}+ v_{H} \right] {\psi^{QP}_{n}} +  \int d\textbf{r' } \Sigma(\textbf{r},\textbf{r'}, \epsilon_{n})\psi^{QP}_{n}(\textbf{r'})  = \epsilon_{n} \psi^{QP}_{n}.
\end{equation}
In calculating our single shot G$_{0}$W$_{0}$ corrections to the Kohn-Sham states we only require the diagonal elements of the self energy. In the Matusbara formalism we can express $\Sigma^{C}(z)$ for points $z=i\nu_{n'}=(2n'+1)\pi i$ along the imaginary axis in terms of the dielectric function at frequencies $i\omega_{n}=2n\pi i$:
\begin{equation}
\Sigma_{nn \textbf{k}}^{C}(i\nu_{n'}) = \frac{iT}{2\pi} \sum_{GG',m,i\omega_{n}}  \frac{[M^{mn}_{G}(\textbf{k})]^{*}M^{mn}_{G'}(\textbf{k}) W_{GG'}^{C}(\omega_{n})}{i\nu_{n'} + i\omega_{n} - \epsilon_{m}^{KS}(\textbf{k})},
\end{equation}
where the Kohn-Sham matrix elements are $M^{mn}_{G}=\matrixel{\psi_{m}^{KS}(\textbf{k})}{e^{iGr}}{\psi_{n}^{KS}(\textbf{k})}$, the screened Coulomb interaction is given by $W^{C}_{GG'}(\omega)=(\epsilon^{-1}_{GG'}(\omega) - 1)V_{GG'}^{C}$, and where we have set the chemical potential $\mu=0$ for convenience. The retarded self energy $\Sigma(\omega+i0^{+})$ is then obtained by analytical continuation from the upper-half plane to the real axis. In our calculation we achieve this numerically by fitting to a Pad\'{e} approximant. In condensed matter systems this method of numerical analytic continuation is normally considered to be a less accurate, but quicker, alternative to the contour deformation method\,\cite{Lebegue2003}. We find it to be sufficient for its primary purpose here of correcting the position of the L-edge.

\begin{figure}
\centering
\includegraphics[clip=false, width=1\linewidth]{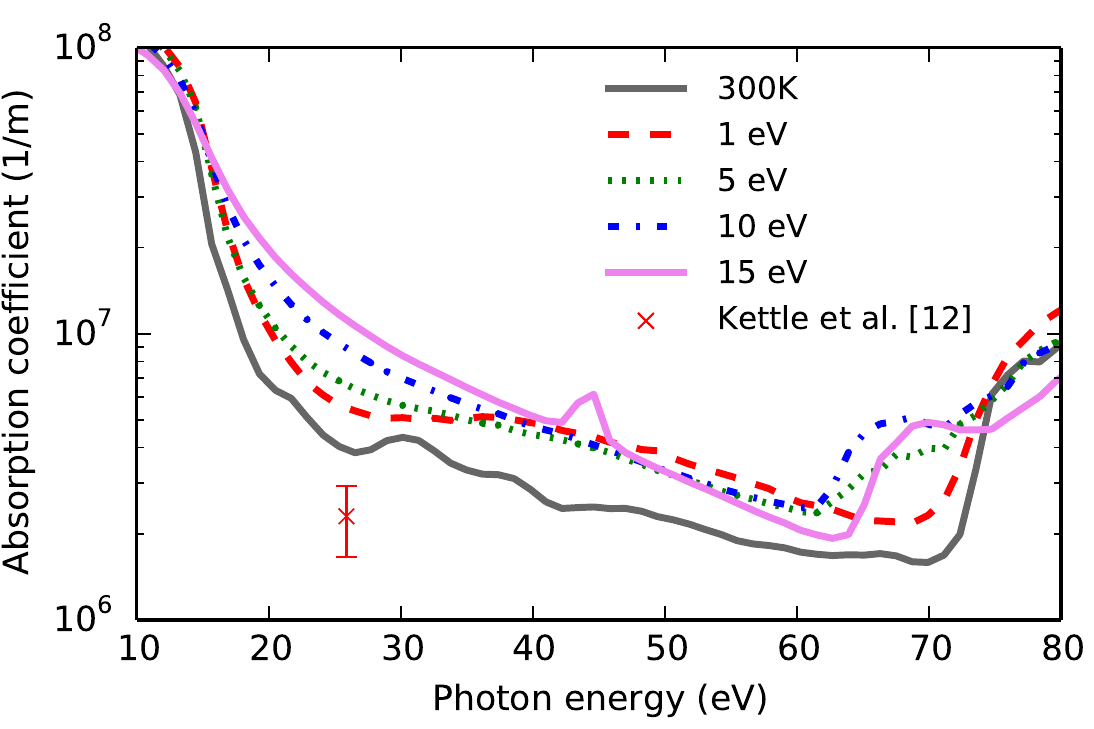}
\caption{(Color online). Calculated opacity for equilibrated warm dense Al up to $T=$ 15\,eV along with the single data point of Kettle et al. ($T_{e}\approx1$\,eV). }
\label{fig:warm}
\end{figure}

We show in Fig.\,\ref{fig:warm} our calculations extended into the warm dense matter regime and plot the predicted opacity for equilibrium, solid density Al at $T= 1, 5, 10$ and $15$\,eV. In moving from room temperature to $T=1$\,eV the free-free opacity increases considerably. This is in contrast to previous average-atom\,\cite{Shaffer2017} and IB calculations\,\cite{Iglesias2010}, as well as being in disagreement with the recent measurement of Kettle et al.\,\cite{Kettle2016}. We note, however, that the cold opacity result from the same authors is also in disagreement with several experimental datasets, and with our theoretical predictions. 

Calculations performed using room temperature ion configurations, but with the electrons heated to 1\,eV, suggest a breakdown of the regular crystalline lattice to be responsible for the sudden increase in free-free absorption at $T=1$\,eV. This may explain why such an increase was not predicted by the previous IB theory (in which the role of ion-ion correlations was only estimated) or average-atom calculations. A key difference between the latter and our calculations is the use here of periodic supercells, well suited to representing the periodic Al FCC crystal structure in the cold limit, and capable of dealing with non-spherically symmetric ion distributions. The discrepancies between these three models suggest the ion distribution, boundary conditions and symmetries of the system may have significant impact on the absorption in warm dense matter at low temperatures. 

This observation of significant changes to the free-free opacity induced by melting is consistent with the recent measurements of Williams et al.\,\cite{Williams2018}, who also attribute their observation to the breakdown of the crystal structure. However, here we predict larger opacity increases than those reported by Williams et al., by about a factor of 2. We note that our calculated opacities are also larger than those previously published by Vinko et al.\,\cite{Vinko2009a} for both cold and warm systems.

For $\hbar\omega<25$\,eV the temperature dependence of the opacity is consistent with plasmon broadening due to increased electron-ion collisions\,\cite{Witte2017},. For $35< \hbar\omega<60$\,eV only a relatively weak temperature dependence is predicted with the exception of the 2s-2p resonance ($\hbar \omega \approx 42$\,eV) owing to thermal ionisation of the L-shell at the highest temperatures. We identify this weak dependence as partly a consequence of LFC effects enhancing the opacity at $T=1$\,eV. The strong impact of LFC at $T=1$\,eV further suggests the importance of ion-ion correlations at this temperature. Close to the L-edge, significant pre-edge features are predicted to develop as low lying continuum states are thermally depopulated. This should be distinguished from changes in the continuum lowering which remains relatively constant, only changing by $\approx4$\,eV for the highest temperature of 15\,eV.

\begin{figure}
\centering
\includegraphics[clip=true, width=1\linewidth]{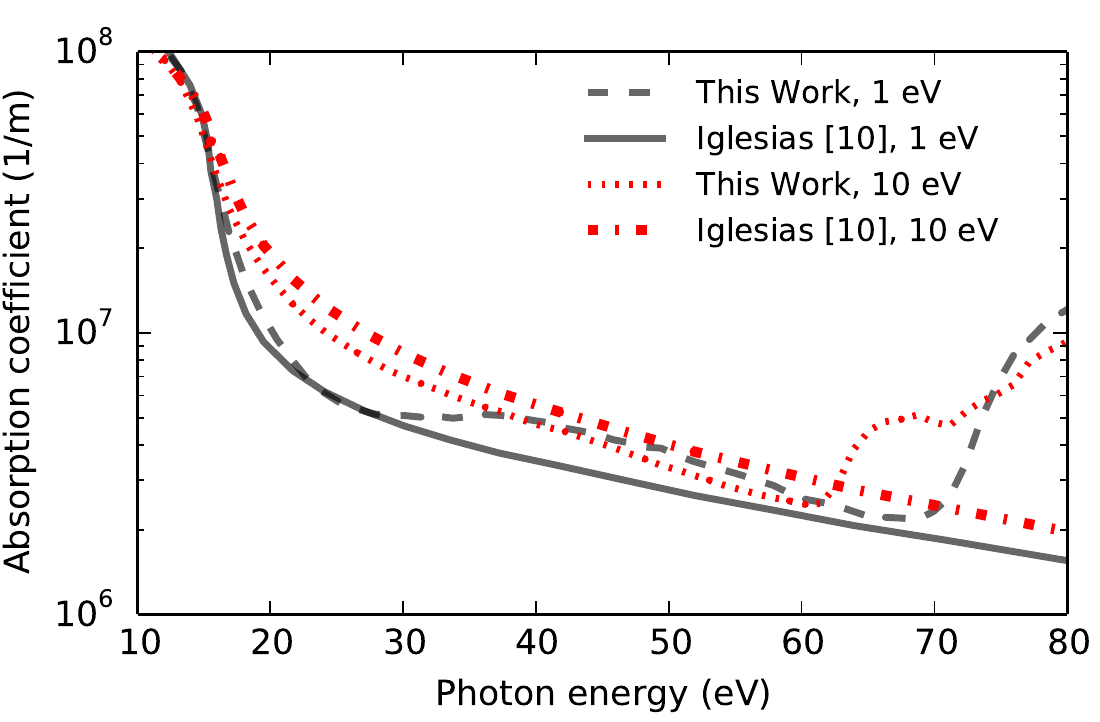}
\caption{(Color online). Comparison of our calculated opacity with the Inverse Bremsstrahlung model of reference\,\cite{Iglesias2010} at high temperatures. Our results appear to approach the IB model both in slope and absolute value.}
\label{fig:IBlim}
 \end{figure}

Large scale DFT, and in particular G$_{0}$W$_{0}$, calculations are expensive at higher temperatures. As such, there is a clear incentive to establish under what conditions simpler models, such as IB calculations, may be sufficiently accurate. In Fig.\,\ref{fig:IBlim} we plot our finite-temperature calculations for 1 and 10\,eV along with the IB model of\,\cite{Iglesias2010}. The latter uses a frozen core pseudopotential and therefore cannot be expected to replicate the 2s-2p bound-bound transition or L-edge features present in our work here. Nonetheless, the IB calculations closely agree with our results for $\hbar\omega<25$\,eV at a temperature of 1\,eV where the opacity is dominated by the plasmon feature, and at higher temperatures our calculations do appear to be approaching the IB opacity model both in slope and absolute values.

In conclusion, we have presented measurements and calculations of the free-free opacity in Al between the plasma frequency and the L-shell absorption edge. Our experimental data shows good agreement with previous measurements by Henke et al.\,\cite{Henke1993} and Birken et al.\,\cite{Birken1986}, and disagrees with the measurements by Kettle et al.\,\cite{Kettle2016}, and with Gullikson et al.\,\cite{Gullikson1994} at lower photon energies.
We find excellent agreement between our measurements and full time-dependent DFT calculations of the free-free opacity. 
We have extended our simulations up to temperatures of 15\,eV, demonstrating and quantifying the influence of finite-temperature G$_{0}$W$_{0}$ corrections in warm dense matter conditions. 
IB-based plasma models show good agreement with TD-DFT methods at temperatures above around 10\,eV. This result is particularly encouraging given the substantial difficulties in conducting full GW-TD-DFT-MD simulations at higher temperatures, and provides confidence in more approximate but faster plasma models for radiative properties of warm dense matter.
Finally, our results demonstrate the importance of finite-temperature quasi-particle and local field corrections in DFT-based modelling, with important implications for the evaluation of the {\it ab-initio} dielectric response more generally, including calculations of dynamic structure factors and stopping powers.

We wish to thank the Target Fabrication Group at the Central Laser Facility, for manufacturing and characterizing our target samples.
P.H., J.S.W and S.M.V. acknowledge support from the U.K. EPSRC under grant EP/P015794/1.
S.M.V. gratefully acknowledges support from the Royal Society.

\bibliographystyle{apsrev4-1}
\bibliography{MyCollection}

\begin{thebibliography}{63}%
\makeatletter
\providecommand \@ifxundefined [1]{%
 \@ifx{#1\undefined}
}%
\providecommand \@ifnum [1]{%
 \ifnum #1\expandafter \@firstoftwo
 \else \expandafter \@secondoftwo
 \fi
}%
\providecommand \@ifx [1]{%
 \ifx #1\expandafter \@firstoftwo
 \else \expandafter \@secondoftwo
 \fi
}%
\providecommand \natexlab [1]{#1}%
\providecommand \enquote  [1]{``#1''}%
\providecommand \bibnamefont  [1]{#1}%
\providecommand \bibfnamefont [1]{#1}%
\providecommand \citenamefont [1]{#1}%
\providecommand \href@noop [0]{\@secondoftwo}%
\providecommand \href [0]{\begingroup \@sanitize@url \@href}%
\providecommand \@href[1]{\@@startlink{#1}\@@href}%
\providecommand \@@href[1]{\endgroup#1\@@endlink}%
\providecommand \@sanitize@url [0]{\catcode `\\12\catcode `\$12\catcode
  `\&12\catcode `\#12\catcode `\^12\catcode `\_12\catcode `\%12\relax}%
\providecommand \@@startlink[1]{}%
\providecommand \@@endlink[0]{}%
\providecommand \url  [0]{\begingroup\@sanitize@url \@url }%
\providecommand \@url [1]{\endgroup\@href {#1}{\urlprefix }}%
\providecommand \urlprefix  [0]{URL }%
\providecommand \Eprint [0]{\href }%
\providecommand \doibase [0]{http://dx.doi.org/}%
\providecommand \selectlanguage [0]{\@gobble}%
\providecommand \bibinfo  [0]{\@secondoftwo}%
\providecommand \bibfield  [0]{\@secondoftwo}%
\providecommand \translation [1]{[#1]}%
\providecommand \BibitemOpen [0]{}%
\providecommand \bibitemStop [0]{}%
\providecommand \bibitemNoStop [0]{.\EOS\space}%
\providecommand \EOS [0]{\spacefactor3000\relax}%
\providecommand \BibitemShut  [1]{\csname bibitem#1\endcsname}%
\let\auto@bib@innerbib\@empty
\bibitem [{\citenamefont {Kramers}(1923)}]{Kramers1923}%
  \BibitemOpen
  \bibfield  {author} {\bibinfo {author} {\bibfnamefont {H.~A.}\ \bibnamefont
  {Kramers}},\ }\href {\doibase 10.1080/14786442308565244} {\bibfield
  {journal} {\bibinfo  {journal} {Philos. Mag.}\ }\textbf {\bibinfo {volume}
  {46}},\ \bibinfo {pages} {836} (\bibinfo {year} {1923})}\BibitemShut
  {NoStop}%
\bibitem [{\citenamefont {Johnston}(1967)}]{Johnston1967}%
  \BibitemOpen
  \bibfield  {author} {\bibinfo {author} {\bibfnamefont {R.~R.}\ \bibnamefont
  {Johnston}},\ }\href@noop {} {\bibfield  {journal} {\bibinfo  {journal} {J.
  Quant. Spectrosc. Radiat. Transfer}\ }\textbf {\bibinfo {volume} {7}},\
  \bibinfo {pages} {815} (\bibinfo {year} {1967})}\BibitemShut {NoStop}%
\bibitem [{\citenamefont {Seely}\ and\ \citenamefont
  {Harris}(1973)}]{Seely1973}%
  \BibitemOpen
  \bibfield  {author} {\bibinfo {author} {\bibfnamefont {J.~F.}\ \bibnamefont
  {Seely}}\ and\ \bibinfo {author} {\bibfnamefont {E.~G.}\ \bibnamefont
  {Harris}},\ }\href {\doibase 10.1103/PhysRevA.7.1064} {\bibfield  {journal}
  {\bibinfo  {journal} {Phys. Rev. A}\ }\textbf {\bibinfo {volume} {7}},\
  \bibinfo {pages} {1064} (\bibinfo {year} {1973})}\BibitemShut {NoStop}%
\bibitem [{\citenamefont {Tsytovich}\ \emph {et~al.}(1995)\citenamefont
  {Tsytovich}, \citenamefont {Bingham}, \citenamefont {de~Angelis},\ and\
  \citenamefont {Forlani}}]{Tsytovich1995}%
  \BibitemOpen
  \bibfield  {author} {\bibinfo {author} {\bibfnamefont {V.~N.}\ \bibnamefont
  {Tsytovich}}, \bibinfo {author} {\bibfnamefont {R.}~\bibnamefont {Bingham}},
  \bibinfo {author} {\bibfnamefont {U.}~\bibnamefont {de~Angelis}}, \ and\
  \bibinfo {author} {\bibfnamefont {A.}~\bibnamefont {Forlani}},\ }\href
  {\doibase 10.1016/0375-9601(95)00549-I} {\bibfield  {journal} {\bibinfo
  {journal} {Phys. Lett. A.}\ }\textbf {\bibinfo {volume} {205}},\ \bibinfo
  {pages} {199} (\bibinfo {year} {1995})}\BibitemShut {NoStop}%
\bibitem [{\citenamefont {Totsuji}(1985)}]{Totsuji1985}%
  \BibitemOpen
  \bibfield  {author} {\bibinfo {author} {\bibfnamefont {H.}~\bibnamefont
  {Totsuji}},\ }\href {\doibase 10.1103/PhysRevA.32.3005} {\bibfield  {journal}
  {\bibinfo  {journal} {Phys. Rev. A}\ }\textbf {\bibinfo {volume} {32}},\
  \bibinfo {pages} {3005} (\bibinfo {year} {1985})}\BibitemShut {NoStop}%
\bibitem [{\citenamefont {Dawson}\ and\ \citenamefont
  {Oberman}(1962)}]{Dawson1962}%
  \BibitemOpen
  \bibfield  {author} {\bibinfo {author} {\bibfnamefont {J.}~\bibnamefont
  {Dawson}}\ and\ \bibinfo {author} {\bibfnamefont {C.}~\bibnamefont
  {Oberman}},\ }\href {\doibase 10.1063/1.1706652} {\bibfield  {journal}
  {\bibinfo  {journal} {Phys. Fluids}\ }\textbf {\bibinfo {volume} {5}},\
  \bibinfo {pages} {517} (\bibinfo {year} {1962})},\ \Eprint
  {http://arxiv.org/abs/arXiv:1011.1669v3} {arXiv:arXiv:1011.1669v3}
  \BibitemShut {NoStop}%
\bibitem [{\citenamefont {Bekefi}(1966)}]{Bekefi1966}%
  \BibitemOpen
  \bibfield  {author} {\bibinfo {author} {\bibfnamefont {G.}~\bibnamefont
  {Bekefi}},\ }\href@noop {} {\emph {\bibinfo {title} {{Radiation Processes in
  Plasmas}}}}\ (\bibinfo  {publisher} {Wiley},\ \bibinfo {year}
  {1966})\BibitemShut {NoStop}%
\bibitem [{\citenamefont {Ron}\ and\ \citenamefont {Tzoar}(1963)}]{Ron1963}%
  \BibitemOpen
  \bibfield  {author} {\bibinfo {author} {\bibfnamefont {A.}~\bibnamefont
  {Ron}}\ and\ \bibinfo {author} {\bibfnamefont {N.}~\bibnamefont {Tzoar}},\
  }\href {\doibase 10.1103/PhysRevLett.10.45} {\bibfield  {journal} {\bibinfo
  {journal} {Phys. Rev.}\ }\textbf {\bibinfo {volume} {131}},\ \bibinfo {pages}
  {12} (\bibinfo {year} {1963})}\BibitemShut {NoStop}%
\bibitem [{\citenamefont {Vinko}\ \emph {et~al.}(2009)\citenamefont {Vinko},
  \citenamefont {Gregori}, \citenamefont {Desjarlais}, \citenamefont {Nagler},
  \citenamefont {Whitcher}, \citenamefont {Lee}, \citenamefont {Audebert},\
  and\ \citenamefont {Wark}}]{Vinko2009a}%
  \BibitemOpen
  \bibfield  {author} {\bibinfo {author} {\bibfnamefont {S.~M.}\ \bibnamefont
  {Vinko}}, \bibinfo {author} {\bibfnamefont {G.}~\bibnamefont {Gregori}},
  \bibinfo {author} {\bibfnamefont {M.~P.}\ \bibnamefont {Desjarlais}},
  \bibinfo {author} {\bibfnamefont {B.}~\bibnamefont {Nagler}}, \bibinfo
  {author} {\bibfnamefont {T.~J.}\ \bibnamefont {Whitcher}}, \bibinfo {author}
  {\bibfnamefont {R.~W.}\ \bibnamefont {Lee}}, \bibinfo {author} {\bibfnamefont
  {P.}~\bibnamefont {Audebert}}, \ and\ \bibinfo {author} {\bibfnamefont
  {J.~S.}\ \bibnamefont {Wark}},\ }\href {\doibase 10.1016/j.hedp.2009.04.004}
  {\bibfield  {journal} {\bibinfo  {journal} {High Energy Density Phys.}\
  }\textbf {\bibinfo {volume} {5}},\ \bibinfo {pages} {124} (\bibinfo {year}
  {2009})}\BibitemShut {NoStop}%
\bibitem [{\citenamefont {Iglesias}(2010)}]{Iglesias2010}%
  \BibitemOpen
  \bibfield  {author} {\bibinfo {author} {\bibfnamefont {C.~A.}\ \bibnamefont
  {Iglesias}},\ }\href {\doibase 10.1016/j.hedp.2010.01.004} {\bibfield
  {journal} {\bibinfo  {journal} {High Energy Density Phys.}\ }\textbf
  {\bibinfo {volume} {6}},\ \bibinfo {pages} {311} (\bibinfo {year}
  {2010})}\BibitemShut {NoStop}%
\bibitem [{\citenamefont {Williams}\ \emph {et~al.}(2013)\citenamefont
  {Williams}, \citenamefont {Chung}, \citenamefont {Vinko}, \citenamefont
  {Künzel}, \citenamefont {Sardinha}, \citenamefont {Zeitoun},\ and\
  \citenamefont {Fajardo}}]{Williams2013}%
  \BibitemOpen
  \bibfield  {author} {\bibinfo {author} {\bibfnamefont {G.~O.}\ \bibnamefont
  {Williams}}, \bibinfo {author} {\bibfnamefont {H.-K.}\ \bibnamefont {Chung}},
  \bibinfo {author} {\bibfnamefont {S.~M.}\ \bibnamefont {Vinko}}, \bibinfo
  {author} {\bibfnamefont {S.}~\bibnamefont {Künzel}}, \bibinfo {author}
  {\bibfnamefont {A.~B.}\ \bibnamefont {Sardinha}}, \bibinfo {author}
  {\bibfnamefont {P.}~\bibnamefont {Zeitoun}}, \ and\ \bibinfo {author}
  {\bibfnamefont {M.}~\bibnamefont {Fajardo}},\ }\href {\doibase
  10.1063/1.4794964} {\bibfield  {journal} {\bibinfo  {journal} {Phys.
  Plasmas}\ }\textbf {\bibinfo {volume} {20}},\ \bibinfo {pages} {042701}
  (\bibinfo {year} {2013})}\BibitemShut {NoStop}%
\bibitem [{\citenamefont {{Kettle $\backslash$emph{\{}et
  al.{\}}}}(2016)}]{Kettle2016}%
  \BibitemOpen
  \bibfield  {author} {\bibinfo {author} {\bibfnamefont {B.}~\bibnamefont
  {{Kettle \emph{et al.}}}},\ }\href {\doibase
  10.1103/PhysRevE.94.023203} {\bibfield  {journal} {\bibinfo  {journal} {Phys.
  Rev. E}\ }\textbf {\bibinfo {volume} {94}},\ \bibinfo {pages} {023203}
  (\bibinfo {year} {2016})}\BibitemShut {NoStop}%
\bibitem [{\citenamefont {Shaffer}\ \emph {et~al.}(2017)\citenamefont
  {Shaffer}, \citenamefont {Ferris}, \citenamefont {Colgan}, \citenamefont
  {Kilcrease},\ and\ \citenamefont {Starrett}}]{Shaffer2017}%
  \BibitemOpen
  \bibfield  {author} {\bibinfo {author} {\bibfnamefont {N.}~\bibnamefont
  {Shaffer}}, \bibinfo {author} {\bibfnamefont {N.}~\bibnamefont {Ferris}},
  \bibinfo {author} {\bibfnamefont {J.}~\bibnamefont {Colgan}}, \bibinfo
  {author} {\bibfnamefont {D.}~\bibnamefont {Kilcrease}}, \ and\ \bibinfo
  {author} {\bibfnamefont {C.}~\bibnamefont {Starrett}},\ }\href {\doibase
  10.1016/j.hedp.2017.02.008} {\bibfield  {journal} {\bibinfo  {journal} {High
  Energy Density Phys.}\ } (\bibinfo {year} {2017}),\
  10.1016/j.hedp.2017.02.008}\BibitemShut {NoStop}%
\bibitem [{\citenamefont {{Williams $\backslash$emph{\{}et
  al.{\}}}}(2018)}]{Williams2018}%
  \BibitemOpen
  \bibfield  {author} {\bibinfo {author} {\bibfnamefont {G.~O.}\ \bibnamefont
  {{Williams \emph{et al.}}}},\ }\href {\doibase
  10.1103/PhysRevA.97.023414} {\bibfield  {journal} {\bibinfo  {journal} {Phys.
  Rev. A}\ }\textbf {\bibinfo {volume} {97}},\ \bibinfo {pages} {023414}
  (\bibinfo {year} {2018})}\BibitemShut {NoStop}%
\bibitem [{\citenamefont {Hopfield}(1965)}]{Hopfield1965}%
  \BibitemOpen
  \bibfield  {author} {\bibinfo {author} {\bibfnamefont {J.~J.}\ \bibnamefont
  {Hopfield}},\ }\href {\doibase 10.1103/PhysRev.139.A419} {\bibfield
  {journal} {\bibinfo  {journal} {Phys. Rev.}\ }\textbf {\bibinfo {volume}
  {139}} (\bibinfo {year} {1965}),\ 10.1103/PhysRev.139.A419}\BibitemShut
  {NoStop}%
\bibitem [{\citenamefont {Sturm}\ and\ \citenamefont
  {Pajanne}(1973)}]{Sturm1973}%
  \BibitemOpen
  \bibfield  {author} {\bibinfo {author} {\bibfnamefont {K.}~\bibnamefont
  {Sturm}}\ and\ \bibinfo {author} {\bibfnamefont {E.}~\bibnamefont
  {Pajanne}},\ }\href {\doibase 10.1088/0305-4608/3/1/032} {\bibfield
  {journal} {\bibinfo  {journal} {J. Phys. F}\ }\textbf {\bibinfo {volume}
  {3}},\ \bibinfo {pages} {199} (\bibinfo {year} {1973})}\BibitemShut {NoStop}%
\bibitem [{\citenamefont {Sturm}(1982)}]{Sturm1982}%
  \BibitemOpen
  \bibfield  {author} {\bibinfo {author} {\bibfnamefont {K.}~\bibnamefont
  {Sturm}},\ }\href {\doibase 10.1080/00018738200101348} {\bibfield  {journal}
  {\bibinfo  {journal} {Adv. Phys.}\ }\textbf {\bibinfo {volume} {31}},\
  \bibinfo {pages} {1} (\bibinfo {year} {1982})}\BibitemShut {NoStop}%
\bibitem [{\citenamefont {Sturm}\ \emph {et~al.}(1990)\citenamefont {Sturm},
  \citenamefont {Zaremba},\ and\ \citenamefont {Nuroh}}]{Sturm1990}%
  \BibitemOpen
  \bibfield  {author} {\bibinfo {author} {\bibfnamefont {K.}~\bibnamefont
  {Sturm}}, \bibinfo {author} {\bibfnamefont {E.}~\bibnamefont {Zaremba}}, \
  and\ \bibinfo {author} {\bibfnamefont {K.}~\bibnamefont {Nuroh}},\ }\href
  {\doibase 10.1103/PhysRevB.42.6973} {\bibfield  {journal} {\bibinfo
  {journal} {Phys. Rev. B}\ }\textbf {\bibinfo {volume} {42}},\ \bibinfo
  {pages} {6973} (\bibinfo {year} {1990})}\BibitemShut {NoStop}%
\bibitem [{\citenamefont {{Zylstra $\backslash$emph{\{}et
  al{\}}}}(2015)}]{Zylstra2015}%
  \BibitemOpen
  \bibfield  {author} {\bibinfo {author} {\bibfnamefont {A.~B.}\ \bibnamefont
  {{Zylstra \emph{et al}}}},\ }\href {\doibase
  10.1103/PhysRevLett.114.215002} {\bibfield  {journal} {\bibinfo  {journal}
  {Phys. Rev. Lett.}\ }\textbf {\bibinfo {volume} {114}},\ \bibinfo {pages} {1}
  (\bibinfo {year} {2015})}\BibitemShut {NoStop}%
\bibitem [{\citenamefont {Roth}\ \emph {et~al.}(2017)\citenamefont {Roth},
  \citenamefont {Bruckner}, \citenamefont {Moro}, \citenamefont {Gruber},
  \citenamefont {Goebl}, \citenamefont {Juaristi}, \citenamefont {Alducin},
  \citenamefont {Steinberger}, \citenamefont {Duchoslav}, \citenamefont
  {Primetzhofer},\ and\ \citenamefont {Bauer}}]{Roth2017}%
  \BibitemOpen
  \bibfield  {author} {\bibinfo {author} {\bibfnamefont {D.}~\bibnamefont
  {Roth}}, \bibinfo {author} {\bibfnamefont {B.}~\bibnamefont {Bruckner}},
  \bibinfo {author} {\bibfnamefont {M.~V.}\ \bibnamefont {Moro}}, \bibinfo
  {author} {\bibfnamefont {S.}~\bibnamefont {Gruber}}, \bibinfo {author}
  {\bibfnamefont {D.}~\bibnamefont {Goebl}}, \bibinfo {author} {\bibfnamefont
  {J.~I.}\ \bibnamefont {Juaristi}}, \bibinfo {author} {\bibfnamefont
  {M.}~\bibnamefont {Alducin}}, \bibinfo {author} {\bibfnamefont
  {R.}~\bibnamefont {Steinberger}}, \bibinfo {author} {\bibfnamefont
  {J.}~\bibnamefont {Duchoslav}}, \bibinfo {author} {\bibfnamefont
  {D.}~\bibnamefont {Primetzhofer}}, \ and\ \bibinfo {author} {\bibfnamefont
  {P.}~\bibnamefont {Bauer}},\ }\href {\doibase 10.1103/PhysRevLett.118.103401}
  {\bibfield  {journal} {\bibinfo  {journal} {Phys. Rev. Lett.}\ }\textbf
  {\bibinfo {volume} {118}},\ \bibinfo {pages} {1} (\bibinfo {year}
  {2017})}\BibitemShut {NoStop}%
\bibitem [{\citenamefont {Baczewski}\ \emph {et~al.}(2016)\citenamefont
  {Baczewski}, \citenamefont {Shulenburger}, \citenamefont {Desjarlais},
  \citenamefont {Hansen},\ and\ \citenamefont {Magyar}}]{Baczewski2016}%
  \BibitemOpen
  \bibfield  {author} {\bibinfo {author} {\bibfnamefont {A.~D.}\ \bibnamefont
  {Baczewski}}, \bibinfo {author} {\bibfnamefont {L.}~\bibnamefont
  {Shulenburger}}, \bibinfo {author} {\bibfnamefont {M.~P.}\ \bibnamefont
  {Desjarlais}}, \bibinfo {author} {\bibfnamefont {S.~B.}\ \bibnamefont
  {Hansen}}, \ and\ \bibinfo {author} {\bibfnamefont {R.~J.}\ \bibnamefont
  {Magyar}},\ }\href {\doibase 10.1103/PhysRevLett.116.115004} {\bibfield
  {journal} {\bibinfo  {journal} {Phys. Rev. Lett.}\ }\textbf {\bibinfo
  {volume} {116}},\ \bibinfo {pages} {115004} (\bibinfo {year}
  {2016})}\BibitemShut {NoStop}%
\bibitem [{\citenamefont {{CXRO Database}}()}]{CXRO}%
  \BibitemOpen
  \bibfield  {author} {\bibinfo {author} {\bibnamefont {{CXRO Database}}},\
  }\href {http://henke.lbl.gov/optical constants/} {\enquote {\bibinfo {title}
  {http://henke.lbl.gov/optical constants/},}\ }\BibitemShut {NoStop}%
\bibitem [{\citenamefont {Gullikson}\ \emph {et~al.}(1994)\citenamefont
  {Gullikson}, \citenamefont {Denham}, \citenamefont {Mrowka},\ and\
  \citenamefont {Underwood}}]{Gullikson1994}%
  \BibitemOpen
  \bibfield  {author} {\bibinfo {author} {\bibfnamefont {E.~M.}\ \bibnamefont
  {Gullikson}}, \bibinfo {author} {\bibfnamefont {P.}~\bibnamefont {Denham}},
  \bibinfo {author} {\bibfnamefont {S.}~\bibnamefont {Mrowka}}, \ and\ \bibinfo
  {author} {\bibfnamefont {J.~H.}\ \bibnamefont {Underwood}},\ }\href@noop {}
  {\bibfield  {journal} {\bibinfo  {journal} {Phys. Rev. B}\ }\textbf {\bibinfo
  {volume} {49}},\ \bibinfo {pages} {16283} (\bibinfo {year}
  {1994})}\BibitemShut {NoStop}%
\bibitem [{\citenamefont {Henke}\ \emph {et~al.}(1993)\citenamefont {Henke},
  \citenamefont {Gullikson},\ and\ \citenamefont {Davis}}]{Henke1993}%
  \BibitemOpen
  \bibfield  {author} {\bibinfo {author} {\bibfnamefont {B.~L.}\ \bibnamefont
  {Henke}}, \bibinfo {author} {\bibfnamefont {E.~M.}\ \bibnamefont
  {Gullikson}}, \ and\ \bibinfo {author} {\bibfnamefont {J.~C.}\ \bibnamefont
  {Davis}},\ }\href {\doibase 10.1006/adnd.1993.1013} {\bibfield  {journal}
  {\bibinfo  {journal} {At. Data Nucl. Data Tables}\ }\textbf {\bibinfo
  {volume} {54}},\ \bibinfo {pages} {181} (\bibinfo {year} {1993})}\BibitemShut
  {NoStop}%
\bibitem [{\citenamefont {Keenan}\ \emph {et~al.}(2002)\citenamefont {Keenan},
  \citenamefont {Lewis}, \citenamefont {Wark},\ and\ \citenamefont
  {Wolfrum}}]{Keenan2002}%
  \BibitemOpen
  \bibfield  {author} {\bibinfo {author} {\bibfnamefont {R.}~\bibnamefont
  {Keenan}}, \bibinfo {author} {\bibfnamefont {C.~L.~S.}\ \bibnamefont
  {Lewis}}, \bibinfo {author} {\bibfnamefont {J.~S.}\ \bibnamefont {Wark}}, \
  and\ \bibinfo {author} {\bibfnamefont {E.}~\bibnamefont {Wolfrum}},\ }\href
  {\doibase 10.1088/0953-4075/35/20/102} {\bibfield  {journal} {\bibinfo
  {journal} {J. Phys. B}\ }\textbf {\bibinfo {volume} {35}},\ \bibinfo {pages}
  {L447} (\bibinfo {year} {2002})}\BibitemShut {NoStop}%
\bibitem [{\citenamefont {Frassetto}\ \emph {et~al.}(2011)\citenamefont
  {Frassetto}, \citenamefont {Cacho}, \citenamefont {Froud}, \citenamefont
  {Turcu}, \citenamefont {Villoresi}, \citenamefont {Bryan}, \citenamefont
  {Springate},\ and\ \citenamefont {Poletto}}]{Frassetto2011}%
  \BibitemOpen
  \bibfield  {author} {\bibinfo {author} {\bibfnamefont {F.}~\bibnamefont
  {Frassetto}}, \bibinfo {author} {\bibfnamefont {C.}~\bibnamefont {Cacho}},
  \bibinfo {author} {\bibfnamefont {C.~A.}\ \bibnamefont {Froud}}, \bibinfo
  {author} {\bibfnamefont {I.~E.}\ \bibnamefont {Turcu}}, \bibinfo {author}
  {\bibfnamefont {P.}~\bibnamefont {Villoresi}}, \bibinfo {author}
  {\bibfnamefont {W.~A.}\ \bibnamefont {Bryan}}, \bibinfo {author}
  {\bibfnamefont {E.}~\bibnamefont {Springate}}, \ and\ \bibinfo {author}
  {\bibfnamefont {L.}~\bibnamefont {Poletto}},\ }\href {\doibase
  10.1364/OE.19.019169} {\bibfield  {journal} {\bibinfo  {journal} {Optics
  Express}\ }\textbf {\bibinfo {volume} {19}},\ \bibinfo {pages} {19169}
  (\bibinfo {year} {2011})}\BibitemShut {NoStop}%
\bibitem [{\citenamefont {Birken}\ \emph {et~al.}(1986)\citenamefont {Birken},
  \citenamefont {Jark}, \citenamefont {Kunz},\ and\ \citenamefont
  {Wolf}}]{Birken1986}%
  \BibitemOpen
  \bibfield  {author} {\bibinfo {author} {\bibfnamefont {H.~G.}\ \bibnamefont
  {Birken}}, \bibinfo {author} {\bibfnamefont {W.}~\bibnamefont {Jark}},
  \bibinfo {author} {\bibfnamefont {C.}~\bibnamefont {Kunz}}, \ and\ \bibinfo
  {author} {\bibfnamefont {R.}~\bibnamefont {Wolf}},\ }\href {\doibase
  10.1016/0168-9002(86)91141-1} {\bibfield  {journal} {\bibinfo  {journal}
  {Nucl. Instrum. Methods}\ }\textbf {\bibinfo {volume} {253}},\ \bibinfo
  {pages} {166} (\bibinfo {year} {1986})}\BibitemShut {NoStop}%
\bibitem [{\citenamefont {Witte}\ \emph {et~al.}(2017)\citenamefont {Witte},
  \citenamefont {Fletcher}, \citenamefont {Galtier}, \citenamefont {Gamboa},
  \citenamefont {Lee}, \citenamefont {Zastrau}, \citenamefont {Redmer},
  \citenamefont {Glenzer},\ and\ \citenamefont {Sperling}}]{Witte2017}%
  \BibitemOpen
  \bibfield  {author} {\bibinfo {author} {\bibfnamefont {B.~B.~L.}\
  \bibnamefont {Witte}}, \bibinfo {author} {\bibfnamefont {L.~B.}\ \bibnamefont
  {Fletcher}}, \bibinfo {author} {\bibfnamefont {E.}~\bibnamefont {Galtier}},
  \bibinfo {author} {\bibfnamefont {E.}~\bibnamefont {Gamboa}}, \bibinfo
  {author} {\bibfnamefont {H.~J.}\ \bibnamefont {Lee}}, \bibinfo {author}
  {\bibfnamefont {U.}~\bibnamefont {Zastrau}}, \bibinfo {author} {\bibfnamefont
  {R.}~\bibnamefont {Redmer}}, \bibinfo {author} {\bibfnamefont {S.~H.}\
  \bibnamefont {Glenzer}}, \ and\ \bibinfo {author} {\bibfnamefont
  {P.}~\bibnamefont {Sperling}},\ }\href {\doibase
  10.1103/PhysRevLett.118.225001} {\bibfield  {journal} {\bibinfo  {journal}
  {Phys. Rev. Lett.}\ }\textbf {\bibinfo {volume} {118}},\ \bibinfo {pages}
  {225001} (\bibinfo {year} {2017})}\BibitemShut {NoStop}%
\bibitem [{\citenamefont {Maitra}(2016)}]{Maitra2016}%
  \BibitemOpen
  \bibfield  {author} {\bibinfo {author} {\bibfnamefont {N.~T.}\ \bibnamefont
  {Maitra}},\ }\href {\doibase 10.1063/1.4953039} {\bibfield  {journal}
  {\bibinfo  {journal} {J. Chem. Phys.}\ }\textbf {\bibinfo {volume} {144}},\
  \bibinfo {pages} {220901} (\bibinfo {year} {2016})}\BibitemShut {NoStop}%
\bibitem [{\citenamefont {van Leeuwen}(1999)}]{VanLeeuwen1999}%
  \BibitemOpen
  \bibfield  {author} {\bibinfo {author} {\bibfnamefont {R.}~\bibnamefont {van
  Leeuwen}},\ }\href {\doibase 10.1103/PhysRevLett.82.3863} {\bibfield
  {journal} {\bibinfo  {journal} {Phys. Rev. Lett.}\ }\textbf {\bibinfo
  {volume} {82}},\ \bibinfo {pages} {3863} (\bibinfo {year}
  {1999})}\BibitemShut {NoStop}%
\bibitem [{\citenamefont {Maitra}\ and\ \citenamefont
  {Burke}(2001)}]{Maitra2001}%
  \BibitemOpen
  \bibfield  {author} {\bibinfo {author} {\bibfnamefont {N.~T.}\ \bibnamefont
  {Maitra}}\ and\ \bibinfo {author} {\bibfnamefont {K.}~\bibnamefont {Burke}},\
  }\href {\doibase 10.1103/PhysRevA.63.042501} {\bibfield  {journal} {\bibinfo
  {journal} {Phys. Rev. A}\ }\textbf {\bibinfo {volume} {63}},\ \bibinfo
  {pages} {042501} (\bibinfo {year} {2001})}\BibitemShut {NoStop}%
\bibitem [{\citenamefont {Dharma-wardana}(2016)}]{Dharma-wardana2016}%
  \BibitemOpen
  \bibfield  {author} {\bibinfo {author} {\bibfnamefont {M.}~\bibnamefont
  {Dharma-wardana}},\ }\href {\doibase 10.3390/computation4020016} {\bibfield
  {journal} {\bibinfo  {journal} {Computation}\ }\textbf {\bibinfo {volume}
  {4}},\ \bibinfo {pages} {16} (\bibinfo {year} {2016})}\BibitemShut {NoStop}%
\bibitem [{\citenamefont {Perrot}\ and\ \citenamefont
  {Dharma-wardana}(1984)}]{Perrot1984}%
  \BibitemOpen
  \bibfield  {author} {\bibinfo {author} {\bibfnamefont {F.}~\bibnamefont
  {Perrot}}\ and\ \bibinfo {author} {\bibfnamefont {M.~W.~C.}\ \bibnamefont
  {Dharma-wardana}},\ }\href {\doibase 10.1103/PhysRevA.30.2619} {\bibfield
  {journal} {\bibinfo  {journal} {Phys. Rev. A}\ }\textbf {\bibinfo {volume}
  {30}},\ \bibinfo {pages} {2619} (\bibinfo {year} {1984})}\BibitemShut
  {NoStop}%
\bibitem [{\citenamefont {Dharma-Wardana}\ and\ \citenamefont
  {Taylor}(1981)}]{Dharma-Wardana1981a}%
  \BibitemOpen
  \bibfield  {author} {\bibinfo {author} {\bibfnamefont {M.~W.}\ \bibnamefont
  {Dharma-Wardana}}\ and\ \bibinfo {author} {\bibfnamefont {R.}~\bibnamefont
  {Taylor}},\ }\href {\doibase 10.1088/0022-3719/14/5/011} {\bibfield
  {journal} {\bibinfo  {journal} {J. Phys. C}\ }\textbf {\bibinfo {volume}
  {14}},\ \bibinfo {pages} {629} (\bibinfo {year} {1981})}\BibitemShut
  {NoStop}%
\bibitem [{\citenamefont {Perrot}\ and\ \citenamefont
  {Dharma-Wardana}(2000)}]{Perrot2000}%
  \BibitemOpen
  \bibfield  {author} {\bibinfo {author} {\bibfnamefont {F.}~\bibnamefont
  {Perrot}}\ and\ \bibinfo {author} {\bibfnamefont {M.~W.~C.}\ \bibnamefont
  {Dharma-Wardana}},\ }\href {\doibase 10.1103/PhysRevB.62.16536} {\bibfield
  {journal} {\bibinfo  {journal} {Phys. Rev. B}\ }\textbf {\bibinfo {volume}
  {62}},\ \bibinfo {pages} {16536} (\bibinfo {year} {2000})}\BibitemShut
  {NoStop}%
\bibitem [{\citenamefont {Pribram-Jones}\ \emph {et~al.}(2016)\citenamefont
  {Pribram-Jones}, \citenamefont {Grabowski},\ and\ \citenamefont
  {Burke}}]{Pribram-Jones2016}%
  \BibitemOpen
  \bibfield  {author} {\bibinfo {author} {\bibfnamefont {A.}~\bibnamefont
  {Pribram-Jones}}, \bibinfo {author} {\bibfnamefont {P.~E.}\ \bibnamefont
  {Grabowski}}, \ and\ \bibinfo {author} {\bibfnamefont {K.}~\bibnamefont
  {Burke}},\ }\href {\doibase 10.1103/PhysRevLett.116.233001} {\bibfield
  {journal} {\bibinfo  {journal} {Phys. Rev. Lett.}\ }\textbf {\bibinfo
  {volume} {116}},\ \bibinfo {pages} {233001} (\bibinfo {year}
  {2016})}\BibitemShut {NoStop}%
\bibitem [{\citenamefont {Hedin}(1965)}]{Hedin1965}%
  \BibitemOpen
  \bibfield  {author} {\bibinfo {author} {\bibfnamefont {L.}~\bibnamefont
  {Hedin}},\ }\href {\doibase 10.1103/PhysRev.139.A796} {\bibfield  {journal}
  {\bibinfo  {journal} {Phys. Rev.}\ }\textbf {\bibinfo {volume} {139}},\
  \bibinfo {pages} {A796} (\bibinfo {year} {1965})},\ \Eprint
  {http://arxiv.org/abs/9712013v1} {arXiv:9712013v1 [arXiv:cond-mat]}
  \BibitemShut {NoStop}%
\bibitem [{\citenamefont {Hedin}(1999)}]{Hedin1999}%
  \BibitemOpen
  \bibfield  {author} {\bibinfo {author} {\bibfnamefont {L.}~\bibnamefont
  {Hedin}},\ }\href@noop {} {\bibfield  {journal} {\bibinfo  {journal} {J.
  Phys. Condens. Matter}\ }\textbf {\bibinfo {volume} {11}},\ \bibinfo {pages}
  {R489} (\bibinfo {year} {1999})}\BibitemShut {NoStop}%
\bibitem [{\citenamefont {Faleev}\ \emph {et~al.}(2006)\citenamefont {Faleev},
  \citenamefont {{Van Schilfgaarde}}, \citenamefont {Kotani}, \citenamefont
  {L{\'{e}}onard},\ and\ \citenamefont {Desjarlais}}]{Faleev2006}%
  \BibitemOpen
  \bibfield  {author} {\bibinfo {author} {\bibfnamefont {S.~V.}\ \bibnamefont
  {Faleev}}, \bibinfo {author} {\bibfnamefont {M.}~\bibnamefont {{Van
  Schilfgaarde}}}, \bibinfo {author} {\bibfnamefont {T.}~\bibnamefont
  {Kotani}}, \bibinfo {author} {\bibfnamefont {F.}~\bibnamefont
  {L{\'{e}}onard}}, \ and\ \bibinfo {author} {\bibfnamefont {M.~P.}\
  \bibnamefont {Desjarlais}},\ }\href {\doibase 10.1103/PhysRevB.74.033101}
  {\bibfield  {journal} {\bibinfo  {journal} {Phys. Rev. B}\ }\textbf {\bibinfo
  {volume} {74}},\ \bibinfo {pages} {033101} (\bibinfo {year}
  {2006})}\BibitemShut {NoStop}%
\bibitem [{\citenamefont {Zimmermann}\ \emph {et~al.}(1978)\citenamefont
  {Zimmermann}, \citenamefont {Kilimann}, \citenamefont {Kraeft}, \citenamefont
  {Kremp},\ and\ \citenamefont {R{\"{o}}pke}}]{Zimmermann1978}%
  \BibitemOpen
  \bibfield  {author} {\bibinfo {author} {\bibfnamefont {R.}~\bibnamefont
  {Zimmermann}}, \bibinfo {author} {\bibfnamefont {K.}~\bibnamefont
  {Kilimann}}, \bibinfo {author} {\bibfnamefont {W.~D.}\ \bibnamefont
  {Kraeft}}, \bibinfo {author} {\bibfnamefont {D.}~\bibnamefont {Kremp}}, \
  and\ \bibinfo {author} {\bibfnamefont {G.}~\bibnamefont {R{\"{o}}pke}},\
  }\href {\doibase 10.1002/pssb.2220900119} {\bibfield  {journal} {\bibinfo
  {journal} {Phys. Status Solidi B}\ }\textbf {\bibinfo {volume} {90}},\
  \bibinfo {pages} {175} (\bibinfo {year} {1978})},\ \Eprint
  {http://arxiv.org/abs/arXiv:1011.1669v3} {arXiv:arXiv:1011.1669v3}
  \BibitemShut {NoStop}%
\bibitem [{\citenamefont {R{\"{o}}pke}\ \emph {et~al.}(1978)\citenamefont
  {R{\"{o}}pke}, \citenamefont {Kilimann}, \citenamefont {Kremp}, \citenamefont
  {Kraeft},\ and\ \citenamefont {Zimmermann}}]{Ropke1978}%
  \BibitemOpen
  \bibfield  {author} {\bibinfo {author} {\bibfnamefont {G.}~\bibnamefont
  {R{\"{o}}pke}}, \bibinfo {author} {\bibfnamefont {K.}~\bibnamefont
  {Kilimann}}, \bibinfo {author} {\bibfnamefont {D.}~\bibnamefont {Kremp}},
  \bibinfo {author} {\bibfnamefont {W.~D.}\ \bibnamefont {Kraeft}}, \ and\
  \bibinfo {author} {\bibfnamefont {R.}~\bibnamefont {Zimmermann}},\ }\href
  {\doibase 10.1002/pssb.2220880158} {\bibfield  {journal} {\bibinfo  {journal}
  {Phys. Status Solidi B}\ }\textbf {\bibinfo {volume} {88}},\ \bibinfo {pages}
  {K59} (\bibinfo {year} {1978})}\BibitemShut {NoStop}%
\bibitem [{\citenamefont {Kraeft}\ \emph {et~al.}(1986)\citenamefont {Kraeft},
  \citenamefont {Kremp}, \citenamefont {Ebeling},\ and\ \citenamefont
  {R{\"{o}}pke}}]{Kraeft1986}%
  \BibitemOpen
  \bibfield  {author} {\bibinfo {author} {\bibfnamefont {W.-D.}\ \bibnamefont
  {Kraeft}}, \bibinfo {author} {\bibfnamefont {D.}~\bibnamefont {Kremp}},
  \bibinfo {author} {\bibfnamefont {W.}~\bibnamefont {Ebeling}}, \ and\
  \bibinfo {author} {\bibfnamefont {G.}~\bibnamefont {R{\"{o}}pke}},\
  }\href@noop {} {\emph {\bibinfo {title} {{Quantum Statistics of Charged
  Particle Systems}}}}\ (\bibinfo  {publisher} {Akademie-Verlag, Berlin, and
  Plenum Press, London/New York},\ \bibinfo {year} {1986})\BibitemShut
  {NoStop}%
\bibitem [{\citenamefont {G{\"{u}}nter}\ \emph {et~al.}(1991)\citenamefont
  {G{\"{u}}nter}, \citenamefont {Hitzschke},\ and\ \citenamefont
  {R{\"{o}}pke}}]{Gunter1991}%
  \BibitemOpen
  \bibfield  {author} {\bibinfo {author} {\bibfnamefont {S.}~\bibnamefont
  {G{\"{u}}nter}}, \bibinfo {author} {\bibfnamefont {L.}~\bibnamefont
  {Hitzschke}}, \ and\ \bibinfo {author} {\bibfnamefont {G.}~\bibnamefont
  {R{\"{o}}pke}},\ }\href {\doibase 10.1103/PhysRevA.44.6834} {\bibfield
  {journal} {\bibinfo  {journal} {Phys. Rev. A}\ }\textbf {\bibinfo {volume}
  {44}},\ \bibinfo {pages} {6834} (\bibinfo {year} {1991})}\BibitemShut
  {NoStop}%
\bibitem [{\citenamefont {Seidel}\ and\ \citenamefont
  {Arndt}(1995)}]{Seidel1995}%
  \BibitemOpen
  \bibfield  {author} {\bibinfo {author} {\bibfnamefont {J.}~\bibnamefont
  {Seidel}}\ and\ \bibinfo {author} {\bibfnamefont {S.}~\bibnamefont {Arndt}},\
  }\href {http://pre.aps.org/abstract/PRE/v52/i5/p5387{\_}1} {\bibfield
  {journal} {\bibinfo  {journal} {Phys. Rev. E}\ }\textbf {\bibinfo {volume}
  {52}},\ \bibinfo {pages} {5387} (\bibinfo {year} {1995})}\BibitemShut
  {NoStop}%
\bibitem [{\citenamefont {Kremp}\ \emph {et~al.}(2005)\citenamefont {Kremp},
  \citenamefont {Schlanges}, \citenamefont {Kraeft},\ and\ \citenamefont
  {Bornath}}]{Kremp2005}%
  \BibitemOpen
  \bibfield  {author} {\bibinfo {author} {\bibfnamefont {D.}~\bibnamefont
  {Kremp}}, \bibinfo {author} {\bibfnamefont {M.}~\bibnamefont {Schlanges}},
  \bibinfo {author} {\bibfnamefont {W.-D.}\ \bibnamefont {Kraeft}}, \ and\
  \bibinfo {author} {\bibfnamefont {T.}~\bibnamefont {Bornath}},\ }\href@noop
  {} {\emph {\bibinfo {title} {{Quantum Statistics of Nonideal Plasmas}}}}\
  (\bibinfo  {publisher} {Springer-Verlag, Berlin/Heidelberg},\ \bibinfo {year}
  {2005})\BibitemShut {NoStop}%
\bibitem [{\citenamefont {Lin}\ \emph {et~al.}(2017)\citenamefont {Lin},
  \citenamefont {R{\"{o}}pke}, \citenamefont {Kraeft},\ and\ \citenamefont
  {Reinholz}}]{Lin2017a}%
  \BibitemOpen
  \bibfield  {author} {\bibinfo {author} {\bibfnamefont {C.}~\bibnamefont
  {Lin}}, \bibinfo {author} {\bibfnamefont {G.}~\bibnamefont {R{\"{o}}pke}},
  \bibinfo {author} {\bibfnamefont {W.~D.}\ \bibnamefont {Kraeft}}, \ and\
  \bibinfo {author} {\bibfnamefont {H.}~\bibnamefont {Reinholz}},\ }\href
  {\doibase 10.1103/PhysRevE.96.013202} {\bibfield  {journal} {\bibinfo
  {journal} {Phys. Rev. E}\ }\textbf {\bibinfo {volume} {96}},\ \bibinfo
  {pages} {013202} (\bibinfo {year} {2017})},\ \Eprint
  {http://arxiv.org/abs/1703.00801} {arXiv:1703.00801} \BibitemShut {NoStop}%
\bibitem [{\citenamefont {Kas}\ and\ \citenamefont {Rehr}(2017)}]{Kas2017}%
  \BibitemOpen
  \bibfield  {author} {\bibinfo {author} {\bibfnamefont {J.~J.}\ \bibnamefont
  {Kas}}\ and\ \bibinfo {author} {\bibfnamefont {J.~J.}\ \bibnamefont {Rehr}},\
  }\href {\doibase 10.1103/PhysRevLett.119.176403} {\bibfield  {journal}
  {\bibinfo  {journal} {Phys. Rev. Lett.}\ }\textbf {\bibinfo {volume} {119}},\
  \bibinfo {pages} {1} (\bibinfo {year} {2017})},\ \Eprint
  {http://arxiv.org/abs/1708.04126} {arXiv:1708.04126} \BibitemShut {NoStop}%
\bibitem [{\citenamefont {Hohenberg}\ and\ \citenamefont {Kohn}(1964)}]{W1964}%
  \BibitemOpen
  \bibfield  {author} {\bibinfo {author} {\bibfnamefont {P.}~\bibnamefont
  {Hohenberg}}\ and\ \bibinfo {author} {\bibfnamefont {W.}~\bibnamefont
  {Kohn}},\ }\href {\doibase 10.1103/PhysRev.136.B864} {\bibfield  {journal}
  {\bibinfo  {journal} {Phys. Rev.}\ }\textbf {\bibinfo {volume} {136}},\
  \bibinfo {pages} {B864} (\bibinfo {year} {1964})}\BibitemShut {NoStop}%
\bibitem [{\citenamefont {Mermin}(1965)}]{Mermin1965}%
  \BibitemOpen
  \bibfield  {author} {\bibinfo {author} {\bibfnamefont {N.~D.}\ \bibnamefont
  {Mermin}},\ }\href
  {http://journals.aps.org/pr/abstract/10.1103/PhysRev.137.A1441} {\bibfield
  {journal} {\bibinfo  {journal} {Phys. Rev.}\ }\textbf {\bibinfo {volume}
  {137}},\ \bibinfo {pages} {1} (\bibinfo {year} {1965})}\BibitemShut {NoStop}%
\bibitem [{\citenamefont {Kohn}\ and\ \citenamefont {Sham}(1965)}]{W1965}%
  \BibitemOpen
  \bibfield  {author} {\bibinfo {author} {\bibfnamefont {W.}~\bibnamefont
  {Kohn}}\ and\ \bibinfo {author} {\bibfnamefont {L.~J.}\ \bibnamefont
  {Sham}},\ }\href {\doibase 10.1103/PhysRev.140.A1133} {\bibfield  {journal}
  {\bibinfo  {journal} {Phys. Rev.}\ }\textbf {\bibinfo {volume} {140}},\
  \bibinfo {pages} {1133} (\bibinfo {year} {1965})}\BibitemShut {NoStop}%
\bibitem [{\citenamefont {Bl{\"{o}}chl}(1994)}]{Blochl1994}%
  \BibitemOpen
  \bibfield  {author} {\bibinfo {author} {\bibfnamefont {P.~E.}\ \bibnamefont
  {Bl{\"{o}}chl}},\ }\href {\doibase 10.1103/PhysRevB.50.17953} {\bibfield
  {journal} {\bibinfo  {journal} {Phys. Rev. B}\ }\textbf {\bibinfo {volume}
  {50}},\ \bibinfo {pages} {17953} (\bibinfo {year} {1994})},\ \Eprint
  {http://arxiv.org/abs/arXiv:1408.4701v2} {arXiv:arXiv:1408.4701v2}
  \BibitemShut {NoStop}%
\bibitem [{\citenamefont {Torrent}\ \emph {et~al.}(2008)\citenamefont
  {Torrent}, \citenamefont {Jollet}, \citenamefont {Bottin}, \citenamefont
  {Z{\'{e}}rah},\ and\ \citenamefont {Gonze}}]{Torrent2008}%
  \BibitemOpen
  \bibfield  {author} {\bibinfo {author} {\bibfnamefont {M.}~\bibnamefont
  {Torrent}}, \bibinfo {author} {\bibfnamefont {F.}~\bibnamefont {Jollet}},
  \bibinfo {author} {\bibfnamefont {F.}~\bibnamefont {Bottin}}, \bibinfo
  {author} {\bibfnamefont {G.}~\bibnamefont {Z{\'{e}}rah}}, \ and\ \bibinfo
  {author} {\bibfnamefont {X.}~\bibnamefont {Gonze}},\ }\href {\doibase
  10.1016/j.commatsci.2007.07.020} {\bibfield  {journal} {\bibinfo  {journal}
  {Comput. Mater. Sci.}\ }\textbf {\bibinfo {volume} {42}},\ \bibinfo {pages}
  {337} (\bibinfo {year} {2008})}\BibitemShut {NoStop}%
\bibitem [{\citenamefont {Runge}\ and\ \citenamefont
  {Gross}(1984)}]{Runge1984}%
  \BibitemOpen
  \bibfield  {author} {\bibinfo {author} {\bibfnamefont {E.}~\bibnamefont
  {Runge}}\ and\ \bibinfo {author} {\bibfnamefont {E.~K.~U.}\ \bibnamefont
  {Gross}},\ }\href {\doibase 10.1103/PhysRevLett.52.997} {\bibfield  {journal}
  {\bibinfo  {journal} {Phys. Rev. Lett.}\ }\textbf {\bibinfo {volume} {52}},\
  \bibinfo {pages} {997} (\bibinfo {year} {1984})}\BibitemShut {NoStop}%
\bibitem [{\citenamefont {Martin}(2010)}]{Martin2010}%
  \BibitemOpen
  \bibfield  {author} {\bibinfo {author} {\bibfnamefont {R.~M.}\ \bibnamefont
  {Martin}},\ }\href@noop {} {\emph {\bibinfo {title} {{Electronic Structure
  Basic Theory and Practical Methods}}}}\ (\bibinfo  {publisher} {Cambridge
  University Press},\ \bibinfo {year} {2010})\BibitemShut {NoStop}%
\bibitem [{\citenamefont {Onida}\ \emph {et~al.}(2002)\citenamefont {Onida},
  \citenamefont {Reining},\ and\ \citenamefont {Rubio}}]{Onida2002}%
  \BibitemOpen
  \bibfield  {author} {\bibinfo {author} {\bibfnamefont {G.}~\bibnamefont
  {Onida}}, \bibinfo {author} {\bibfnamefont {L.}~\bibnamefont {Reining}}, \
  and\ \bibinfo {author} {\bibfnamefont {A.}~\bibnamefont {Rubio}},\ }\href
  {\doibase 10.1103/RevModPhys.74.601} {\bibfield  {journal} {\bibinfo
  {journal} {Rev. Mod. Phys.}\ }\textbf {\bibinfo {volume} {74}},\ \bibinfo
  {pages} {601} (\bibinfo {year} {2002})}\BibitemShut {NoStop}%
\bibitem [{\citenamefont {Desjarlais}\ \emph {et~al.}(2002)\citenamefont
  {Desjarlais}, \citenamefont {Kress},\ and\ \citenamefont
  {Collins}}]{Desjarlais2002}%
  \BibitemOpen
  \bibfield  {author} {\bibinfo {author} {\bibfnamefont {M.~P.}\ \bibnamefont
  {Desjarlais}}, \bibinfo {author} {\bibfnamefont {J.~D.}\ \bibnamefont
  {Kress}}, \ and\ \bibinfo {author} {\bibfnamefont {L.~A.}\ \bibnamefont
  {Collins}},\ }\href {\doibase 10.1103/PhysRevE.66.025401} {\bibfield
  {journal} {\bibinfo  {journal} {Phys. Rev. E}\ }\textbf {\bibinfo {volume}
  {66}},\ \bibinfo {pages} {025401} (\bibinfo {year} {2002})}\BibitemShut
  {NoStop}%
\bibitem [{\citenamefont {Mazevet}\ \emph
  {et~al.}(2005{\natexlab{a}})\citenamefont {Mazevet}, \citenamefont
  {Desjarlais}, \citenamefont {Collins}, \citenamefont {Kress},\ and\
  \citenamefont {Magee}}]{Mazevet2005}%
  \BibitemOpen
  \bibfield  {author} {\bibinfo {author} {\bibfnamefont {S.}~\bibnamefont
  {Mazevet}}, \bibinfo {author} {\bibfnamefont {M.~P.}\ \bibnamefont
  {Desjarlais}}, \bibinfo {author} {\bibfnamefont {L.~A.}\ \bibnamefont
  {Collins}}, \bibinfo {author} {\bibfnamefont {J.~D.}\ \bibnamefont {Kress}},
  \ and\ \bibinfo {author} {\bibfnamefont {N.~H.}\ \bibnamefont {Magee}},\
  }\href {\doibase 10.1103/PhysRevE.71.016409} {\bibfield  {journal} {\bibinfo
  {journal} {Phys. Rev. E}\ }\textbf {\bibinfo {volume} {71}},\ \bibinfo
  {pages} {016409} (\bibinfo {year} {2005}{\natexlab{a}})}\BibitemShut
  {NoStop}%
\bibitem [{\citenamefont {Mazevet}\ \emph
  {et~al.}(2005{\natexlab{b}})\citenamefont {Mazevet}, \citenamefont
  {Cl{\'{e}}rouin}, \citenamefont {Recoules}, \citenamefont {Anglade},\ and\
  \citenamefont {Zerah}}]{Mazevet2005a}%
  \BibitemOpen
  \bibfield  {author} {\bibinfo {author} {\bibfnamefont {S.}~\bibnamefont
  {Mazevet}}, \bibinfo {author} {\bibfnamefont {J.}~\bibnamefont
  {Cl{\'{e}}rouin}}, \bibinfo {author} {\bibfnamefont {V.}~\bibnamefont
  {Recoules}}, \bibinfo {author} {\bibfnamefont {P.~M.}\ \bibnamefont
  {Anglade}}, \ and\ \bibinfo {author} {\bibfnamefont {G.}~\bibnamefont
  {Zerah}},\ }\href {\doibase 10.1103/PhysRevLett.95.085002} {\bibfield
  {journal} {\bibinfo  {journal} {Phys. Rev. Lett.}\ }\textbf {\bibinfo
  {volume} {95}},\ \bibinfo {pages} {085002} (\bibinfo {year}
  {2005}{\natexlab{b}})}\BibitemShut {NoStop}%
\bibitem [{\citenamefont {{Abinit Website}}()}]{AbinitWebsite}%
  \BibitemOpen
  \bibfield  {author} {\bibinfo {author} {\bibnamefont {{Abinit Website}}},\
  }\href {http://www.abinit.org} {\enquote {\bibinfo {title}
  {http://www.abinit.org},}\ }\BibitemShut {NoStop}%
\bibitem [{\citenamefont {{Gonze $\backslash$emph{\{}et
  al.{\}}}}(2016)}]{Gonze2016a}%
  \BibitemOpen
  \bibfield  {author} {\bibinfo {author} {\bibfnamefont {X.}~\bibnamefont
  {{Gonze \emph{et al.}}}},\ }\href {\doibase
  10.1016/j.cpc.2016.04.003} {\bibfield  {journal} {\bibinfo  {journal}
  {Comput. Phys. Commun.}\ }\textbf {\bibinfo {volume} {205}},\ \bibinfo
  {pages} {106} (\bibinfo {year} {2016})}\BibitemShut {NoStop}%
\bibitem [{\citenamefont {Miyake}\ and\ \citenamefont
  {Aryasetiawan}(2000)}]{Miyake2000}%
  \BibitemOpen
  \bibfield  {author} {\bibinfo {author} {\bibfnamefont {T.}~\bibnamefont
  {Miyake}}\ and\ \bibinfo {author} {\bibfnamefont {F.}~\bibnamefont
  {Aryasetiawan}},\ }\href {\doibase 10.1103/PhysRevB.61.7172} {\bibfield
  {journal} {\bibinfo  {journal} {Phys. Rev. B}\ }\textbf {\bibinfo {volume}
  {61}},\ \bibinfo {pages} {7172} (\bibinfo {year} {2000})}\BibitemShut
  {NoStop}%
\bibitem [{\citenamefont {Shishkin}\ and\ \citenamefont
  {Kresse}(2006)}]{Shishkin2006}%
  \BibitemOpen
  \bibfield  {author} {\bibinfo {author} {\bibfnamefont {M.}~\bibnamefont
  {Shishkin}}\ and\ \bibinfo {author} {\bibfnamefont {G.}~\bibnamefont
  {Kresse}},\ }\href {\doibase 10.1103/PhysRevB.74.035101} {\bibfield
  {journal} {\bibinfo  {journal} {Phys. Rev. B}\ }\textbf {\bibinfo {volume}
  {74}},\ \bibinfo {pages} {035101} (\bibinfo {year} {2006})}\BibitemShut
  {NoStop}%
\bibitem [{\citenamefont {Leb{\`{e}}gue}\ \emph {et~al.}(2003)\citenamefont
  {Leb{\`{e}}gue}, \citenamefont {Arnaud}, \citenamefont {Alouani},\ and\
  \citenamefont {Bloechl}}]{Lebegue2003}%
  \BibitemOpen
  \bibfield  {author} {\bibinfo {author} {\bibfnamefont {S.}~\bibnamefont
  {Leb{\`{e}}gue}}, \bibinfo {author} {\bibfnamefont {B.}~\bibnamefont
  {Arnaud}}, \bibinfo {author} {\bibfnamefont {M.}~\bibnamefont {Alouani}}, \
  and\ \bibinfo {author} {\bibfnamefont {P.~E.}\ \bibnamefont {Bloechl}},\
  }\href {\doibase 10.1103/PhysRevB.67.155208} {\bibfield  {journal} {\bibinfo
  {journal} {Phys. Rev. B}\ }\textbf {\bibinfo {volume} {67}},\ \bibinfo
  {pages} {155208} (\bibinfo {year} {2003})},\ \Eprint
  {http://arxiv.org/abs/0301320} {arXiv:0301320 [cond-mat]} \BibitemShut
  {NoStop}%
\end{thebibliography}%

\end{document}